\documentclass[11pt]{article}
\usepackage{amsmath}
\usepackage{mathrsfs}
\usepackage{amsfonts, amssymb, mathrsfs, txfonts}
\usepackage{graphicx, subfigure}
\usepackage{wasysym}
\usepackage{color}

\usepackage{pifont}
\usepackage{mathrsfs}
\usepackage{amsfonts}
\usepackage{graphicx}
\usepackage{amssymb}
\usepackage{txfonts}
\usepackage{amsmath}
\usepackage{amssymb,amsfonts,txfonts, psfrag,eepic,epsfig,graphicx}

\makeatletter \@addtoreset{equation}{section}
\@addtoreset{theo}{section} \makeatother
\renewcommand{\theequation}{\arabic{section}.\arabic{equation}}

\makeatletter
\newfont{\footsc}{cmcsc10 at 8truept}
\newfont{\footbf}{cmbx10 at 8truept}
\newfont{\footrm}{cmr10 at 10truept}
\title{\bf{Conservation laws, bright matter wave solitons and modulational instability
of nonlinear Schr\"{o}dinger equation with time-dependent
nonlinearity} \footnote{Corresponding author. Tel: +86-411-84708351-8136\protect \\
\hspace*{3ex} E-mail address:
shoufu2006@126.com, shoufu@math.ubc.ca (S. F. Tian)} }

\author{ Shou-Fu Tian$^{1,2*}$, Li Zou$^{3}$, Qi Ding$^{1}$ and Hong-Qing Zhang$^{1}$. \\
 \small $^{1}$School of Mathematical Sciences, Dalian University of
Technology, Dalian 116024,\\\small People's Republic of China\\
\small $^{2}$Department of Mathematics, University of British
Columbia, Vancouver, BC, V6T 1Z2, Canada\\
\small $^{3}$School of Aeronautics and Astronautics, Dalian University of Technology, Dalian 116024,
\\\small People's Republic of China}

\date{}
\begin{document}
\maketitle

\noindent {\large \bf Abstract:} In this paper, we consider a
general form of nonlinear Schr\"{o}dinger equation with
time-dependent nonlinearity.  Based on the linear eigenvalue
problem, the complete integrability of such nonlinear Schr\"{o}dinger equation is identified by admitting an infinite
number of conservation laws. Using the Darboux transformation
method, we obtain some explicit bright multi-soliton solutions in a recursive manner. The propagation characteristic of solitons and
their interactions under the periodic plane wave background are
discussed. Finally, the modulational
instability of solutions is analyzed in the presence of small perturbation. 

$~~~~~~~~~~~~~~~~~~~$(Some figures in this article are in colour
only in the electronic version)

\section{Introduction}

It is known that
every atom in a Bose-Einstein Condensates (BECs) moves in an effective mean field due to the
other atoms and the mean field equation of motion governing the
evolution of the macroscopic wave function of the Bose-Einstein
condensate is the so-called time-dependent Gross-Pitaevskii (GP) equation
\cite{Gross}
\begin{equation}
i\hbar\frac{\partial \Psi(\vec{r},t)}{\partial
t}=\left[-\frac{\hbar^{2}\nabla^{2}}{2m}+V_{ext}(\vec{r})+g|
\Psi(\vec{r},t)|^{2}\right] \Psi(\vec{r},t),
\end{equation}
where $\Psi$ is the BEC order parameter, $V_{ext}$ is the external
trapping potential and the coefficient $g=4\pi\hbar^{2}a/m$
characterizes the effective interatomic interactions in the BEC
through the s-wave scattering length $a$. For a parabolic potential
and time dependent scattering lengths $V_{ext}=-\frac{\epsilonup^{2}x^{2}}{4}\hbar$ with $a(t)=-\frac{2\pi \hbar a}{m}$,
$\hbar=2m$, the above GP equation in one
dimension takes the following nonlinear Schr\"{o}dinger equation
with time-dependent nonlinearity
[2-5]
\begin{equation}\label{1D NLSE}
i\frac{\partial \psi(x,t)}{\partial
t}+\frac{\partial^{2}\psi(x,t)}{\partial
x^{2}}+2a(t)|\psi(x,t)|^{2}\psi(x,t)
+\frac{1}{4}\epsilonup^{2}x^{2}\psi(x,t)=0.
\end{equation}
Here the Feshbach-managed nonlinear coefficient $a(t)$ can be redefined as
$a(t)=|a_{s}(t)|/a_{B}=g_{0}\exp(\epsilonup t)$ ($a_{B}$ is the Bohr
radius)\cite{Perez-Garcia}, which is also called the time dependent
scattering length. In Eq.\eqref{1D NLSE}, time $t$ and coordinate
$x$ are measured in units $2/\omega_{\bot}$ and $a_{\bot}$, where
$a_{\bot}=\left(\hbar/m\omega_{\bot}\right)^{1/2}$ and
$a_{0}=\left(\hbar/m\omega_{0}\right)^{1/2}$ are linear oscillator
lengths in the transverse and cigar-axis directions, respectively.
$\omega_{\bot}$ and $\omega_{0}$ are corresponding harmonic
oscillator frequencies, $m$ is the atomic mass, and the parameter
$\epsilonup=2|\omega_{0}|/\omega_{\bot}\ll 1$.

Investigation of the behaviour of
Bose-Einstein Condensates (BECs) requires solving an inhomogeneous
nonlinear Schr\"{o}dinger equation known as the GP equation \cite{Gross}. Eventhough numerical solutions of GP
equation are available \cite{GFT-1999,A-2000}, construction of analytic solutions will
offer more insight into the domain of BECs opening the doors for
developing concrete applications of BECs in future. As we well know, it is significantly important in mathematical physics to search for
exact solutions to equation \eqref{1D NLSE}. Exact solutions
play a vital role in understanding various qualitative and
quantitative features of nonlinear phenomena. It is well known
that searching for soliton solutions of the nonlinear evolution
equations is one of the most important topics in soliton theory.
 Darboux
transformation (DT) \cite{Matveev,Gu} has been proven to be one of the
most fruitful algorithmic procedures to obtain exact solutions of the
nonlinear evolution equations.

The main aim of the present paper is to  construct some infinite
number of conservation laws, explicit bright multi-soliton solutions
by using DT method, and investigate the modulational instability of
solutions of a general form of nonlinear Schr\"{o}dinger equation
with time-dependent nonlinearity \eqref{1D NLSE}.   In this paper,
 on the basis of the Lax pair associated with Eq.\eqref{1D NLSE}, we
will derive an infinite number of conservation laws to identify its
complete integrability. Furthermore, we will apply the Darboux
transformation method to this integrable model and give the general
procedure to recursively generate the bright $N$-soliton solutions
from an initial trivial solution. Moreover, we will discuss the
propagation characteristic and interactions of solitons under
periodic plane wave background and analyze the linear stability of
the nonlinear plane waves.

 The paper will be organized as follows:
In Section 2, based on the linear eigenvalue problem associated with
Eq.\eqref{1D NLSE} and an auxiliary 1D NLSE \eqref{New 1D NLSE}, we
obtain a series of conservation laws. Then the integrability is
identified by admitting an infinite number of conservation laws. In
Section 3, we will apply the Darboux transformation method to this
integrable model and give the general procedure to recursively
generate the bright $N$-soliton solutions from an initial trivial
solution. The propagation characteristic of solitons and their
interactions under the periodic plane wave background are discussed
in Section 4. In Section 5, we analyze the modulational instability
of the nonlinear plane waves. Finally, some conclusions and
discussions are provided.

\section{Lax pair and  infinite  conservation laws}
In this section,  using the linear eigenvalue problem associated
with 1D NLSE \eqref{1D NLSE} and an auxiliary 1D NLSE \eqref{New 1D
NLSE}, we construct  an infinite number of conservation laws for the
1D NLSE \eqref{1D NLSE}. Then the integrability is identified by
admitting an infinite number of conservation laws.

Under the following transformation
\begin{equation}
\psi(x,t)=u(x,t)e^{-i\frac{\epsilonup x^{2}}{4}-\frac{\epsilonup
t}{2}}
\end{equation}
to Eq.\eqref{1D NLSE}, we can obtain the following new 1D NLSE
\begin{equation} \label{New 1D NLSE}
i\frac{\partial}{\partial t}u(x,t)+\frac{\partial^{2}}{\partial
x^{2}}u(x,t)-i\epsilonup x\frac{\partial}{\partial
x}u(x,t)+2a(t)|u(x,t)|^{2}u(x,t)e^{-\epsilonup t}-i \epsilonup
u(x,t)=0.
\end{equation}
By virtue of the Ablowitz-Kaup-Newell-Segur scheme \cite{Ablowitz,AKNS-1974}, the Lax pair
associated with equation \eqref{New 1D NLSE} can be derived as
\begin{equation} \label{lax pair}
\Phi_{x}=U\Phi,\quad \Phi_{t}=V\Phi,
\end{equation}
where $\Phi=(\phiup_{1},\phiup_{2})^{T}$ (the superscript $T$
denotes the vector transpose) is the vector eigenfunction\cite{LZL}, and the
matrices $U$ and $V$ have the following forms
\addtocounter{equation}{1}
\begin{align}
U&=\lambda J+P, \tag{\theequation a}\\
V&=2i\lambda^{2}J+ \epsilonup  \lambda x J P+2i\epsilonup P+Q,
\tag{\theequation b}
\end{align}
with \begin{align*} &J=\left( \begin {array}{cc}
1&0\\\noalign{\medskip}0&-1\end {array} \right),\quad P=\left(
\begin {array}{cc} 0&\sqrt{g_{0}}u(x,t)\\\noalign{\medskip}-\sqrt{g_{0}}u(x,t)^{*}&0\end {array}
\right),\notag\\
&Q=\left(
\begin {array}{cc} ig_{0}|u(x,t)|^{2}&-\sqrt{g_{0}}\epsilonup x u(x,t)+i\sqrt{g_{0}}u(x,t)_{x}
\\\noalign{\medskip}\sqrt{g_{0}}\epsilonup x u(x,t)^{*}+i\sqrt{g_{0}}u(x,t)_{x}^{*}&-ig_{0}|u(x,t)|^{2}\end {array}
\right),
\end{align*}
where $g_{0}=a(t)\exp(-\epsilonup t)$ is an arbitrary function.
 Hereafter the asterisk stands for the complex conjugate. From
the compatibility condition $U_{t}-V_{x}+[U,V]=0$, one can derive
Eq.\eqref{New 1D NLSE} and Eq.\eqref{1D NLSE} in the case of
$\psi(x,t)=u(x,t)e^{-i\frac{\epsilonup x^{2}}{4}-\frac{\epsilonup
t}{2}}$, respectively.

Using Lax pair \eqref{lax pair} and Refs.\cite{R-2000,Tianbo-2010}, we can further derive an
infinite number of conservation laws of Eqs.\eqref{1D NLSE} and
\eqref{New 1D NLSE}. By introducing the quantity
$f(x,t)=\sqrt{g_{0}}u(x,t)\frac{\phiup_{2}}{\phiup_{1}}$, the linear
equations \eqref{lax pair} can be transformed into the following Ricatti
equation
\begin{equation} \label{Ricatti
equation}
f(x,t)_{x}+i\frac{u(x,t)_{xx}}{u(x,t)}f(x,t)+\alpha(x,t)\frac{u(x,t)_{x}}{u(x,t)}f(x,t)-2ig_{0}f(x,t)^{2}
-2ig_{0}u(x,t)u(x,t)^{*}+\beta(x,t)
f(x,t)=0,
\end{equation}
with $\alpha(x,t)=2i\epsilonup-2i+\epsilonup \lambda x-\epsilonup
 x+i\lambda$ and
$\beta(x,t)=\epsilonup\lambda-\epsilonup+\epsilonup\lambda^{2}x+2i\epsilonup\lambda-\epsilonup\lambda
x-4i\lambda^{2} $. Substituting
$f(x,t)=\sum_{n=1}^{\infty}\frac{f_{n}}{\beta(x,t)^{n}}$ into
Ricatti equation \eqref{Ricatti equation} and equating the like
powers of $\beta(x,t)$ to zero, we have a recursion formula
\begin{equation}
f_{1}=2ig_{0}uu^{*},\quad
f_{n+1}=2ig_{0}\sum_{i=1}^{n}f_{i}f_{n-i}-f_{nx}
-i\frac{u_{xx}}{u}f_{n}-\alpha\frac{u_{x}}{u}f_{n}=0,\quad
(n=2,3,\cdots),
\end{equation}
where $f_{n}$ ($n=1,2,\cdots$) are the functions to be determined.
By virtue of the compatibility condition $\left(\ln
\phiup_{1}\right)_{xt}=\left(\ln \phiup_{1}\right)_{tx}$, we obtain
the following conservation form
\begin{equation}
i\frac{\partial}{\partial t}\rho_{k}(x,t)+\frac{\partial}{\partial
x}\mathfrak{I}_{k}(x,t)=0,
\end{equation}
where $\rho_{k}(x,t)$ and $\mathfrak{I}_{k}(x,t)$ ($k=1,2,\cdots$)
are called conserved densities and conserved fluxes, respectively.
The first three significant physical conservation laws are presented
as
\begin{align*}
&\rho_{1}(x,t)=2ig_{0}|u|^{2},\quad
\rho_{2}(x,t)=-2ig_{0}(1+\alpha)u_{x}u^{*}-2ig_{0}uu_{x}^{*}+2g_{0}u_{xx}u^{*},\notag\\
&\rho_{3}(x,t)=-8ig_{0}^{3}|u|^{4}-2g_{0}u_{xx}u_{x}^{*}+2i(1+\alpha)u_{xx}u^{*}+2i\epsilonup
g_{0}(\lambda-1)u_{x}u^{*}+2i(1+3\alpha)g_{0}u_{x}u_{x}^{*}\notag\\
&~~~~~~~~~~~\quad-\frac{1}{u}\left[ 2ig_{0}u_{xx}^{2}u^{*}-2i\alpha
(1+\alpha)g_{0}u_{x}^{2}u^{*}+2(1+2\alpha)g_{0}u_{x}u_{xx}u^{*}\right],\notag\\
&\mathfrak{I}_{1}(x,t)=ig_{0}|u|^{2}-2g_{0}\sqrt{g_{0}}u_{x}u^{*}+2i(\epsilonup\lambda
x+2i\epsilonup -\epsilonup x)g_{0}\sqrt{g_{0}}uu^{*},\notag\\
&\mathfrak{I}_{2}(x,t)=ig_{0}|u|^{2}+2g_{0}\sqrt{g_{0}}u_{x}u_{x}^{*}+2ig_{0}\sqrt{g_{0}}\frac{u_{x}u_{xx}u^{*}}{u}
+2(1+\alpha)g_{0}\sqrt{g_{0}}\frac{u_{x}^{2}u^{*}}{u}
-2ig_{0}\sqrt{g_{0}}(\epsilonup\lambda x+2i\epsilonup\notag\\
&~~~~~~~~~~~\quad-\epsilonup
x)\left[uu_{x}^{*}+iu_{xx}u^{*}+(1+\alpha)u_{x}u^{*}\right],\notag\\
&\mathfrak{I}_{3}(x,t)=ig_{0}|u|^{2}+\sqrt{g_{0}}\left(\epsilonup\lambda
 x+2i\epsilonup-\epsilonup x+i\frac{u_{x}}{u}\right)\notag\\
&~~~~~~~~~~~~~~~~\times \left[-8ig_{0}^{3}|u|^{4}-2g_{0}
 u_{xx}u_{x}^{*}+2i(1+\alpha)u_{xx}u^{*}+2i\epsilonup
 (\lambda-1)g_{0}u_{x}u^{*}+2i(1+3\alpha)g_{0}u_{x}u_{x}^{*}\right]\notag\\
 &~~~~~~~~~~~~~~~~-\frac{\sqrt{g_{0}}}{u}\left(\epsilonup\lambda
 x+2i\epsilonup-\epsilonup x+i\frac{u_{x}}{u}\right)\left[ 2ig_{0}u_{xx}^{2}u^{*}-2i\alpha
(1+\alpha)g_{0}u_{x}^{2}u^{*}+2(1+2\alpha)g_{0}u_{x}u_{xx}u^{*}\right].
\end{align*}
The conserved quantities
$\mathscr{J}=2ig_{0}\int_{-\infty}^{+\infty}|u|^{2}dx$,
$\mathscr{H}=-2ig_{0}\int_{-\infty}^{+\infty}(1+\alpha)u_{x}u^{*}+uu_{x}^{*}+iu_{xx}u^{*}dx$
and
$\mathscr{K}=2ig_{0}\int_{-\infty}^{\infty}-4g_{0}^{2}|u|^{4}+iu_{xx}u_{x}^{*}+\frac{1}{g_{0}}(1+\alpha)u_{xx}u^{*}+\epsilonup
(\lambda-1)u_{x}u^{*}+(1+3\alpha)u_{x}u_{x}^{*}-\frac{1}{u}\left[
u_{xx}^{2}u^{*}-\alpha (1+\alpha)u_{x}^{2}u^{*}\right.$
$\left.-i(1+2\alpha)u_{x}u_{xx}u^{*}\right]dx$ represent the energy,
momentum and Hamiltonian, respectively.

\section{Darboux transformation and bright soliton
solutions} In this section, the Darboux transformation
method is applied to this integrable model and the general procedure is presented to
recursively generate the bright $N$-soliton solutions from an
initial trivial solution. Based on Lax pair \eqref{lax pair}, the
use of the Darboux transformation to construct the bright soliton
solution of nonlinear partial differential equations is an optimum
choice among various methods in soliton theory [15-26]. Owing to its
purely algebraic algorithm, with the symbolic computation, the
analytical $N$-soliton solution can be generated through successive
application of the Darboux transformation [31,32].

\subsection{Darboux transformation}
The Darboux transformation is actually a gauge transformation
$\Phi[1]=T\Phi$ of the spectral problem \eqref{lax pair} by considering
the following gauge transformation
\begin{equation} \label{DT gauge}
\Phi[1]=(\lambda I-S)\Phi~~\mbox{with}~~ S=H\Lambda H^{-1},~~
\Lambda=\rm{diag}(\lambda_{1},\lambda_{1}^{*}),
\end{equation}
where $H$ is a nonsingular matrix. It is required that $\Phi[1]$
solves the same spectral problems \eqref{lax pair}
\begin{equation} \label{DT lax pair}
\Phi[1]_{x}=U[1]\Phi[1]~~\mbox{and}~~ \Phi[1]_{t}=V[1]\Phi[1],
\end{equation}
with $ U[1]=\lambda J+P[1],  V[1]=2i\lambda^{2}J+ \epsilonup \lambda
x J P[1]+2i\epsilonup P[1]+Q[1]$ and
\begin{align}
& P[1]=\left(
\begin {array}{cc} 0&\sqrt{g_{0}}u[1]\\\noalign{\medskip}-\sqrt{g_{0}}u[1]^{*}&0\end {array}
\right),~~J=\left( \begin {array}{cc}
1&0\\\noalign{\medskip}0&-1\end {array} \right),\notag\\
&Q[1]=\left(
\begin {array}{cc} ig_{0}|u[1]|^{2}&-\sqrt{g_{0}}\epsilonup x u[1]+i\sqrt{g_{0}}u[1]_{x}
\\\noalign{\medskip}\sqrt{g_{0}}\epsilonup x u[1]^{*}+i\sqrt{g_{0}}u[1]_{x}^{*}&-ig_{0}|u[1]|^{2}\end {array}
\right),
\end{align}
where $u[1]=\psi[1]e^{i\frac{\epsilonup x^{2}}{4}+\frac{\epsilonup
t}{2}}$. Substituting Eq. \eqref{DT gauge} into Eqs. \eqref{DT
lax pair}, we have the following relationship
\begin{equation} \label{relationship}
P[1]=P+[J,S], \quad S_{x}+[S,JS+P]=0.
\end{equation}
Now we discuss a concrete transformation. It is easy to verify that
if $\Phi=(\phiup_{1},\phiup_{2})^{T}$ is an eigenfunction of
Eqs. \eqref{lax pair} with eigenvalue $\lambda=\lambda_{1}$,
then $(\phiup_{2},-\phiup_{1}^{*})^{T}$ is also an eigenfunction of Eqs. \eqref{lax pair} with eigenvalue $\lambda=\lambda_{1}^{*}$.
 Thus we take the matrix $H$
in the form
\begin{equation} \label{the form H}
H=\left( \begin {array}{cc}
\phiup_{1}&\phiup_{2}^{*}\\\noalign{\medskip}\phiup_{2}&-\phiup_{1}^{*}\end
{array} \right).
\end{equation}
Therefore, by means of Eqs. \eqref{relationship} and
\eqref{the form H}, the onceiterated new potential of Eqs.\eqref{1D
NLSE} and \eqref{New 1D NLSE} are given by
 \begin{align} \label{DT1}
&\psi[1]=u[1]e^{-i\frac{\epsilonup x^{2}}{4}-\frac{\epsilonup
t}{2}},\notag\\
&u[1]=u+2\frac{(\lambda_{1}-\lambda_{1}^{*})\phiup_{1}
\phiup_{2}^{*}}{\phiup_{1}\phiup_{1}^{*}+\phiup_{2}\phiup_{2}^{*}}.
\end{align}
It is straightforward to verify that the Darboux transformations
\eqref{DT gauge} and \eqref{DT1} can simultaneously preserve the
form of linear eigenvalue problem \eqref{lax pair}.

According to the above computation, we have the following
propositions.\\

\noindent \textbf{Proposition 3.1.}\emph{ When $u(x,t)$ is given, let
$(\phiup_{1},\phiup_{2})^{T}$ be the solution of Eqs.\eqref{lax
pair} with $\lambda=\lambda_{k}$, then by using of the Darboux
matrix $T$ \eqref{DT gauge} and Darboux transformation \eqref{DT1},
we have \begin{equation} U[1]=(T_{x}+TU)T^{-1},\quad with\quad
T=\lambda I-S.
\end{equation}}
We can obtain the same proposition about the auxiliary spectral
problem.\\

\noindent \textbf{Proposition 3.2.} \emph{When $u(x,t)$ is given, let
$(\phiup_{1},\phiup_{2})^{T}$ be the solution of Eqs.\eqref{lax
pair} with $\lambda=\lambda_{k}$, then by using of the Darboux
matrix $T$ \eqref{DT gauge} and Darboux transformation \eqref{DT1},
we have \begin{equation} \label{V[1]}
 V[1]=(T_{t}+TV)T^{-1}~~\mbox{with}~~
T=\lambda I-S.
\end{equation}}
\textbf{Proof.}
 Substituting Eqs. \eqref{DT gauge}, \eqref{DT lax pair} and \eqref{DT1}
 into Eq. \eqref{V[1]} by a direct calculation,
we can obtain the conclusion. The proof is
completed.$~~~~~~~~~~~~~~~~~~~~~~~~~~~~~~~~~~~~~~~~~~~~~~~~~~~~~~~~
~~~~~~~~~~~~~~~~~~~~~~~~~~~~~~~~~~~~~~~~~~~$ $\Box$

Propositions 3.1 and 3.2 show that the transformation  \eqref{DT
gauge} and \eqref{DT1} change the Lax pair \eqref{lax pair} into
another Lax pair of the type \eqref{DT lax pair} with $U[1]$ and
$V[1]$ having the same form as $U$ and $V$, respectively. Therefore
both of the Lax pairs lead to the same equation \eqref{New 1D NLSE},
so Eqs. \eqref{DT gauge} and \eqref{DT1} are the Darboux
transformation of Eq.\eqref{New 1D NLSE}. From propositions 3.1 and
3.2, we have following
theorem.\\

\noindent \textbf{Theorem 3.3.}\emph{ Assuming that $\psi,u,
(\phiup[1,\lambda_{1}]_{1},\phiup[1,\lambda_{1}]_{2})^{T},
(\phiup[2,\lambda_{2}]_{1},\phiup[2,\lambda_{2}]_{2})^{T},\cdots,
(\phiup[N,\lambda_{N}]_{1},\phiup[N,\lambda_{N}]_{2})^{T}$ be the
solution of the 1D NLSE \eqref{New 1D NLSE} and $N$ linearly
independent solutions of the linear eigenvalue problem \eqref{lax
pair}, respectively, and after iterating the Darboux transformation
\eqref{DT gauge} and \eqref{DT1} $N$ times analogous to the above
procedure, we can further obtain the $N$th-iterated potential
transformation as
\begin{align} \label{DTN}
&\psi[N]=u[N]e^{-i\frac{\epsilonup x^{2}}{4}-\frac{\epsilonup
t}{2}},\notag\\
&u[N]=u+2\sum_{i=1}^{N}\frac{(\lambda_{i}-\lambda_{i}^{*})\phiup[i,\lambda_{i}]_{1}
\phiup[i,\lambda_{i}]_{2}^{*}}{\phiup[i,\lambda_{i}]_{1}\phiup[i,\lambda_{i}]_{1}^{*}
+\phiup[i,\lambda_{i}]_{2}\phiup[i,\lambda_{i}]_{2}^{*}},
\end{align}
with
\begin{align}\label{AB}
&\phiup[i+1,\lambda_{i+1}]_{j}=(\lambda_{i+1}-\lambda_{i}^{*})\phiup[i,\lambda_{i+1}]_{j}
-\frac{\mathscr{A}_{i}}{\mathscr{B}_{i}}(\lambda_{i}-\lambda_{i}^{*})\phiup[i,\lambda_{i}]_{j},\notag\\
&\mathscr{A}_{i}=\phiup[i,\lambda_{i}]_{1}^{*}\phiup[i,\lambda_{i+1}]_{1}
+\phiup[i,\lambda_{i}]_{2}^{*}\phiup[i,\lambda_{i+1}]_{2},\notag\\
&
\mathscr{B}_{i}=\phiup[i,\lambda_{i+1}]_{1}\phiup[i,\lambda_{i+1}]_{1}^{*}
+\phiup[i,\lambda_{i+1}]_{2}\phiup[i,\lambda_{i+1}]_{2}^{*},~~
(i=1,2,\cdots,N-1,j=1,2),
\end{align}
where
$(\phiup[k,\lambda_{k}]_{1},\phiup[k,\lambda_{k}]_{2}^{*})^{T}$
$(k=1,2,\cdots,N)$ is the eigenfunction of Eqs. \eqref{lax
pair} with the eigenvalue $\lambda=\lambda_{k}$ and potential
$u[k-1]$}.\\
\textbf{Proof.} (By induction) The case $N=1$ follows from Darboux
transformation \eqref{DT gauge} and \eqref{DT1}. Now assume that
equation \eqref{DTN} holds for $N=n$. Then
\begin{align} \label{DTn}
&\psi[n]=u[n]e^{-i\frac{\epsilonup x^{2}}{4}-\frac{\epsilonup
t}{2}},\notag\\
&u[n]=u+2\sum_{i=1}^{n}\frac{(\lambda_{i}-\lambda_{i}^{*})\phiup[i,\lambda_{i}]_{1}
\phiup[i,\lambda_{i}]_{2}^{*}}{\phiup[i,\lambda_{i}]_{1}\phiup[i,\lambda_{i}]_{1}^{*}
+\phiup[i,\lambda_{i}]_{2}\phiup[i,\lambda_{i}]_{2}^{*}},
\end{align}
where
\begin{align}\label{ABn}
&\phiup[i+1,\lambda_{i+1}]_{j}=(\lambda_{i+1}-\lambda_{i}^{*})\phiup[i,\lambda_{i+1}]_{j}
-\frac{\mathscr{A}_{i}}{\mathscr{B}_{i}}(\lambda_{i}-\lambda_{i}^{*})\phiup[i,\lambda_{i}]_{j},\notag\\
&\mathscr{A}_{i}=\phiup[i,\lambda_{i}]_{1}^{*}\phiup[i,\lambda_{i+1}]_{1}
+\phiup[i,\lambda_{i}]_{2}^{*}\phiup[i,\lambda_{i+1}]_{2},\notag\\
&
\mathscr{B}_{i}=\phiup[i,\lambda_{i+1}]_{1}\phiup[i,\lambda_{i+1}]_{1}^{*}
+\phiup[i,\lambda_{i+1}]_{2}\phiup[i,\lambda_{i+1}]_{2}^{*},~~
(i=1,2,\cdots,n-1,j=1,2).
\end{align}
 In case of $N=n+1$, using Darboux transformation \eqref{DT1}, one obtains
\begin{align}\label{DTn+1-1}
u[n+1]=&u[n]+2\frac{(\lambda_{n+1}-\lambda_{n+1}^{*})\phiup[n+1,\lambdaup_{n+1}]_{1}
\phiup[n+1,\lambdaup_{n+1}]_{2}^{*}}{\phiup[n+1,\lambdaup_{n+1}]_{1}\phiup[n+1,\lambdaup_{n+1}]_{1}^{*}
+\phiup[n+1,\lambdaup_{n+1}]_{2}\phiup[n+1,\lambdaup_{n+1}]_{2}^{*}},\notag\\
=&u+2\sum_{i=1}^{n+1}\frac{(\lambda_{i}-\lambda_{i}^{*})\phiup[i,\lambda_{i}]_{1}
\phiup[i,\lambda_{i}]_{2}^{*}}{\phiup[i,\lambda_{i}]_{1}\phiup[i,\lambda_{i}]_{1}^{*}
+\phiup[i,\lambda_{i}]_{2}\phiup[i,\lambda_{i}]_{2}^{*}}.~~(\mbox{with}~Eq.\eqref{DTn})
\end{align}
Based on the first equation of Darboux transformation \eqref{DT1}, it
is easy to obtain
\begin{equation}\label{DTn+1-2}
\psi[n+1]=u[n+1]e^{-i\frac{\epsilonup x^{2}}{4}-\frac{\epsilonup
t}{2}}.
\end{equation}
From Eqs.\eqref{DTn+1-1} and \eqref{DTn+1-2}, the conclusions of
Eqs.\eqref{DTN} and \eqref{AB} are hold for $N=n+1$. This completes
the
proof.$~~~~~~~~~~~~~~~~~~~~~~~~~~~~~~~~~~~~~~~~~~~~~~~~~~~~~~~~~~~~~~~~~~~~~
~~~~~~~~~~~~~~~~~~~~~~~~~~~~~~~~~~~~~~~~~~~~~~~~~~~~~~~~~~~~~~$
$\Box$

\subsection{Bright  matter wave  soliton solution}

Bright soliton here is defined as the emergence of a positive pulse.
Bright soliton solutions have been investigated by original  Refs.\cite{Ablowitz,K-1998,G-1983}.
In the following,  the Darboux transformation is applied to
construct the explicit bright soliton solutions of Eq.\eqref{1D
NLSE}. Using the trivial solution $\psi=0$ (u=0), we solve
the linear equations \eqref{lax pair} with
$\lambda=\lambda_{1}=\frac{1}{2}(\mu_{1}+i\nu_{1})$ and obtain the
eigenfunction
\begin{equation}\label{c1}
\phiup_{1}=e^{\frac{1}{2}(\wp_{1}+i\vartheta_{1})},~~
\phiup_{2}=e^{-\frac{1}{2}(\wp_{1}+i\vartheta_{1})},
\end{equation}
with
\begin{equation}\label{c2}
\wp_{1}=\mu_{1}x-2\mu_{1}\nu_{1}t\xiup_{1},~~
\vartheta_{1}=\nu_{1}x+\left(\mu_{1}^{2}-\nu_{1}^{2}\right)t+\etaup_{1},
\end{equation}
where $\xiup_{1}$ and $\etaup_{1}$ are arbitrary  constants.

 The bright
one-soliton solution for Eq.\eqref{1D NLSE} are obtained as
\begin{align} \label{bright
one-soliton} &\psi(x,t)=u(x,t)e^{-i\frac{\epsilonup
x^{2}}{4}-\frac{\epsilonup t}{2}},\notag\\
&u(x,t)=i\nu_{1}e^{i\vartheta_{1}}\rm{sech}\wp_{1}.
\end{align}
by substitution of the above results into formula \eqref{DT1}.
It implies that the imaginary part
$\nu_{1}$ of the eigenvalue $\lambda_{1}$ determines the amplitude
of the solitons $\psi(x,t)$ and $u(x,t)$, while the velocity of
solitons are related to both real and imaginary parts of the
eigenvalue $\lambda_{1}$.

From above, we have the following propositions.\\

\noindent \textbf{Proposition 3.4.} \emph{Assuming that $
(\phiup[1,\lambda_{1}]_{1},\phiup[1,\lambda_{1}]_{2})^{T}$ and
$(\phiup[2,\lambda_{2}]_{1},\phiup[2,\lambda_{2}]_{2})^{T}$ be two
linearly independent solutions  of the linear eigenvalue problem
\eqref{lax pair} corresponding to two different eigenvalues
$\lambda_{1}=\frac{1}{2}(\mu_{1}+i\nu_{1})$ and
$\lambda_{2}=\frac{1}{2}(\mu_{2}+i\nu_{2})$, respectively, we obtain
the bright two-soliton solutions of Eq.\eqref{1D NLSE} and
Eq.\eqref{New 1D NLSE} from formula \eqref{DTN} with $N=2$
\begin{align} \label{bright two-soliton}
&\psi[2]=u[2]e^{-i\frac{\epsilonup x^{2}}{4}-\frac{\epsilonup
t}{2}},\notag\\
&u[2]=i\frac{\mathscr{U}_{1}e^{i\vartheta_{2}}\rm{cosh}\wp_{1}+
\mathscr{U}_{2}e^{i\vartheta_{1}}\rm{cosh}\wp_{2}+\mathscr{U}_{3}\left(e^{i\vartheta_{1}}\rm{sinh}\wp_{2}
-e^{i\vartheta_{2}}\rm{sinh\wp_{1}}\right)}{\mathscr{V}_{1}\rm{cosh}(\wp_{1}+\wp_{2})+
\mathscr{V}_{2}\rm{cosh}(\wp_{1}-\wp_{2})-2\nu_{1}\nu_{2}\rm{cosh}(\vartheta_{2}-\vartheta_{1})},
\end{align}
with
\begin{align}
&\wp_{i}=\mu_{i}x-2\mu_{i}\nu_{i}t\xiup_{i},~~
\vartheta_{i}=\nu_{i}x+\left(\mu_{i}^{2}-\nu_{i}^{2}\right)t+\etaup_{i}, \notag\\
&\mathscr{U}_{1}=\nu_{2}\left[(\mu_{1}-\mu_{2})^{2}-\nu_{1}^{2}+\nu_{2}^{2}\right],~~
\mathscr{U}_{2}=\nu_{1}\left[(\mu_{1}-\mu_{2})^{2}+\nu_{1}^{2}-\nu_{2}^{2}\right],
\notag\\
&\mathscr{U}_{3}=-2i\nu_{1}\nu_{2}(\mu_{1}-\mu_{2}),~~\mathscr{V}_{1}=\frac{1}{2}(\mu_{1}-\mu_{2})^{2}+\frac{1}{2}(\nu_{1}-\nu_{2})^{2},
\notag\\&\mathscr{V}_{2}=\frac{1}{2}(\mu_{1}-\mu_{2})^{2}+\frac{1}{2}(\nu_{1}+\nu_{2})^{2},~~(i=1,2),
\end{align}
where $\xiup_{i}, \etaup_{i}$ are all complex constants and the
parameters $\mu_{i},\nu_{i}\neq 0 (i=1,2).$}\\
\textbf{Proof.} It is straightforward to prove this proposition by using
the initial values \eqref{c1}, \eqref{c2}, and Theorem 3.3 for
$u=0$, $N=2$.
$~~~~~~~~~~~~~~~~~~~~~~~~~~~~~~~~~~~~~~~~~~~~~~~~~~~~~~~~~~~~~~~~~~~~~~~~~~~~~~~~~~~~~~~~~~~~~~
~~~~~~~~~~~~~~~~~~~~~~~~~~~~~$$\Box$

\noindent \textbf{Proposition 3.5.} \emph{Assuming that $
(\phiup[1,\lambda_{1}]_{1},\phiup[1,\lambda_{1}]_{2})^{T},
(\phiup[2,\lambda_{2}]_{1},\phiup[2,\lambda_{2}]_{2})^{T},\cdots,
(\phiup[N,\lambda_{N}]_{1},\phiup[N,\lambda_{N}]_{2})^{T}$ be $N$
linearly independent solutions  of the linear eigenvalue problem
\eqref{lax pair} corresponding to $N$ different eigenvalues
$\lambda_{1}=\frac{1}{2}(\mu_{1}+i\nu_{1}),\lambda_{2}=\frac{1}{2}(\mu_{2}+i\nu_{2}),\cdots,
\lambda_{N}=\frac{1}{2}(\mu_{N}+i\nu_{N})$, respectively, we obtain
the bright $N$-soliton solutions of Eqs.\eqref{1D NLSE} and
\eqref{New 1D NLSE} from formula \eqref{DTN} with $N$
\begin{align} \label{DTNsolutions}
&\psi[N]=u[N]e^{-i\frac{\epsilonup x^{2}}{4}-\frac{\epsilonup
t}{2}},\notag\\
&u[N]=2\sum_{i=1}^{N}\frac{(\lambda_{i}-\lambda_{i}^{*})\phiup[i,\lambda_{i}]_{1}
\phiup[i,\lambda_{i}]_{2}^{*}}{\phiup[i,\lambda_{i}]_{1}\phiup[i,\lambda_{i}]_{1}^{*}
+\phiup[i,\lambda_{i}]_{2}\phiup[i,\lambda_{i}]_{2}^{*}},
\end{align}
with  \begin{equation}
\wp_{i}=\mu_{i}x-2\mu_{i}\nu_{i}t\xiup_{i},~~
\vartheta_{i}=\nu_{i}x+\left(\mu_{i}^{2}-\nu_{i}^{2}\right)t+\etaup_{i},\quad
(i=1,2,\cdots,N),
\end{equation} where  $\xiup_{i}, \etaup_{i}$ are all real constants and the
parameters $\mu_{i},\nu_{i}\neq 0 (i=1,2,\cdots,N).$
$(\phiup[k,\lambda_{k}]_{1},\phiup[k,\lambda_{k}]_{2}^{*})^{T}$
satisfies Eq.\eqref{AB} and is the eigenfunction of Eqs.
\eqref{lax pair} with the eigenvalue $\lambda=\lambda_{k}$ and
potential $u[k]$ $(k=1,2,\cdots,N)$.} \\
\textbf{Proof.} It is straightforward to prove this proposition by using
the initial values \eqref{c1}, \eqref{c2}, and Theorem 3.3 for
$u=0$.
$~~~~~~~~~~~~~~~~~~~~~~~~~~~~~~~~~~~~~~~~~~~~~~~~~~~~~~~~~~~~~~~~~~~~~~~~~~~~~~~
~~~~~~~~~~~~~~~~~~~~~~~~~~~~~~~~~~~~~~~~~$$\Box$

The graphs of the bright one-soliton periodic wave solution
\eqref{bright one-soliton} and two-soliton periodic wave solution
\eqref{bright two-soliton} are plotted in Fig. 1 and Fig. 2,
respectively.\\

$~~~~~~~~$
{\rotatebox{0}{\includegraphics[width=5cm,height=4cm,angle=0]{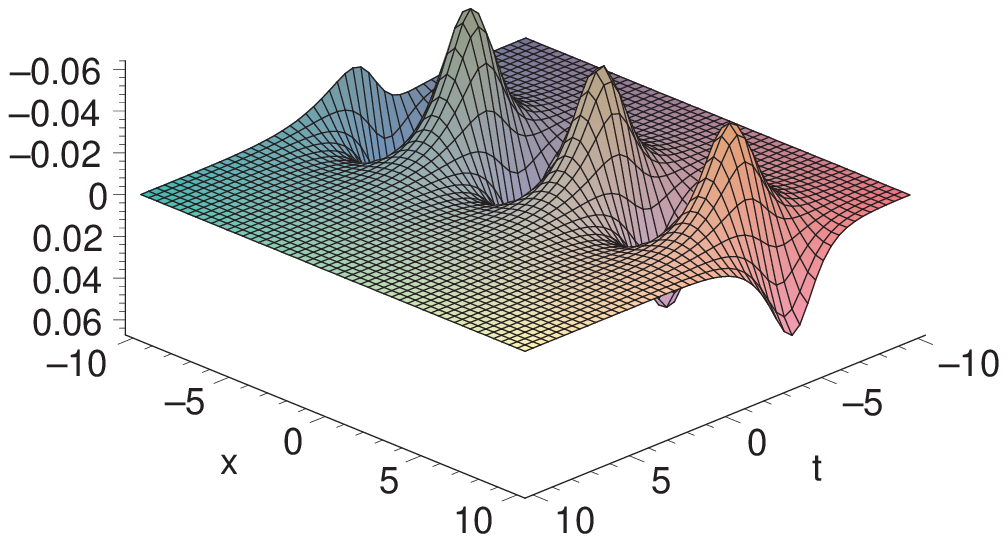}}}\qquad
$~~~~~~~~~~~~~$
{\rotatebox{0}{\includegraphics[width=5cm,height=4cm,angle=0]{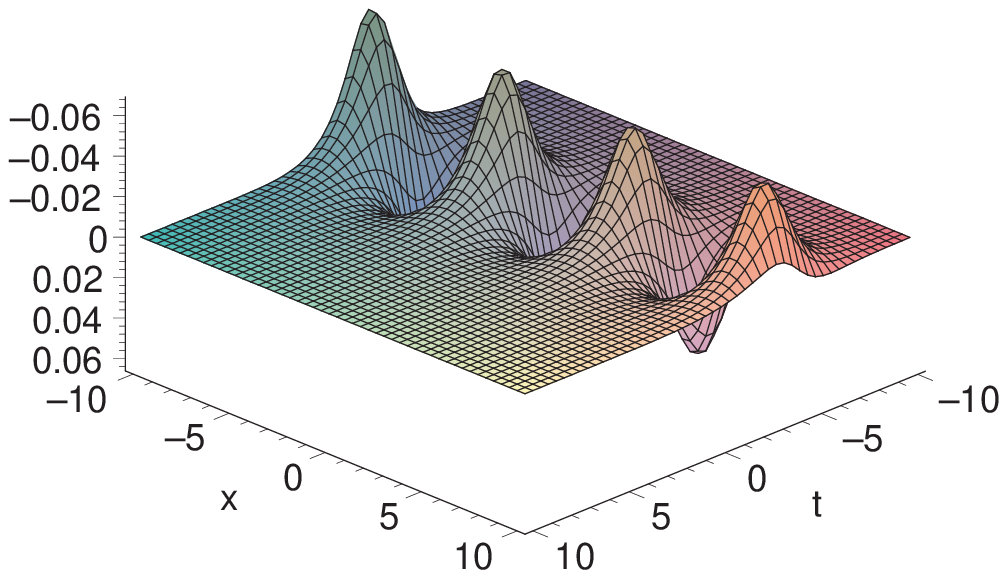}}}%\quad\\

$~~~~~~~~~~~~~~~~~~~~~~~~~~~~~~~~~~$$(a)$$
~~~~~~~~~~~~~~~~~~~~~~~~~~~~~~~~~~~~~~~~~~~~~~~~~~~~~~~~~~~~~~~~~~~~~~$
 $(b)$

 $~~~~$
{\rotatebox{0}{\includegraphics[width=5cm,height=4cm,angle=0]{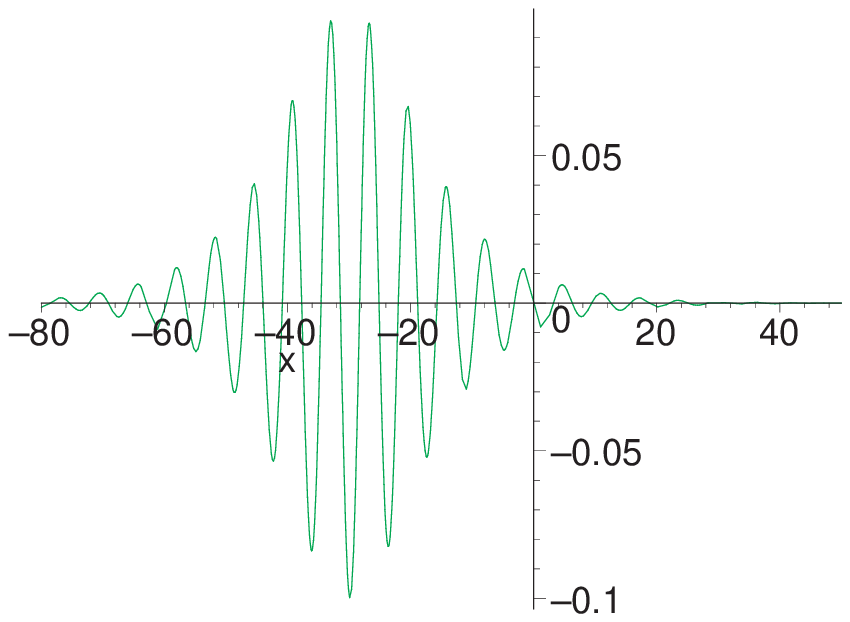}}}\qquad
$~~~~~~~~~~~~~$
{\rotatebox{0}{\includegraphics[width=5cm,height=4cm,angle=0]{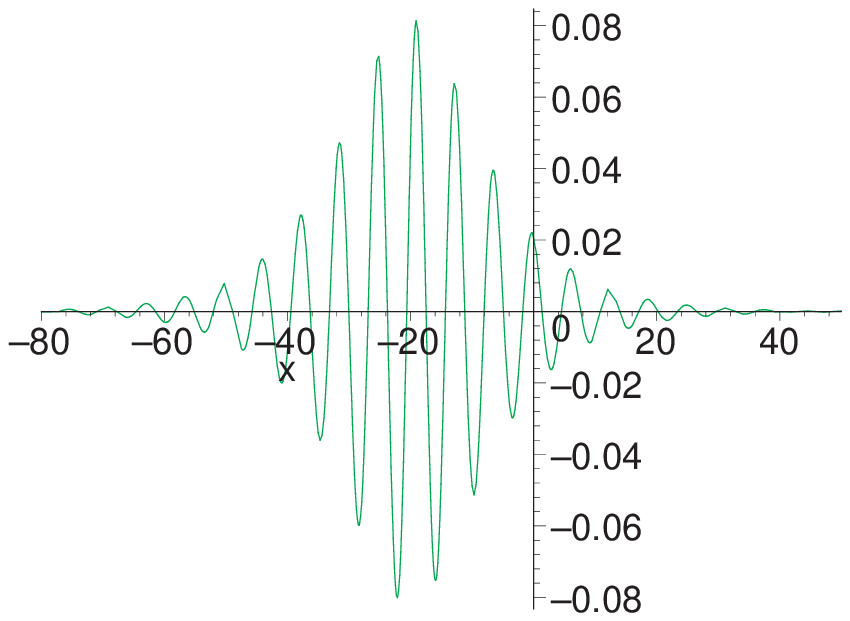}}}%\quad\\

$~~~~~~~~~~~~~~~~~~~~~~~~~~~~~~~~~~$$(c)$$
~~~~~~~~~~~~~~~~~~~~~~~~~~~~~~~~~~~~~~~~~~~~~~~~~~~~~~~~~~~~~~~~~~~~~~$
 $(d)$\\
\small{\textbf{Fig. 1.} (Color online) A  symmetric bright
one-soliton solution $|\psi(x,t)|^{2}$ of Eq.\eqref{1D NLSE} with
parameters: $\mu_{1}=1$, $\nu_{1}=0.1$, $\xiup_{1}=1+i$,
$\etaup_{1}=3$ and  $\epsilonup=0.01$. This figure shows that the
symmetric bright one-soliton periodic wave is spatially periodic in
two directions, but it need not be periodic in either the $x$ or $t$
directions. $(a)$ Perspective view for the real part of wave. $(b)$
Perspective view of the imaginary part wave. $(c)$ Wave propagation
pattern of the real part wave along the $x$ axis. $(d)$ Wave
propagation pattern of the imaginary part wave along the $x$ axis.}

$~~~~~~~~$
{\rotatebox{0}{\includegraphics[width=5cm,height=4cm,angle=0]{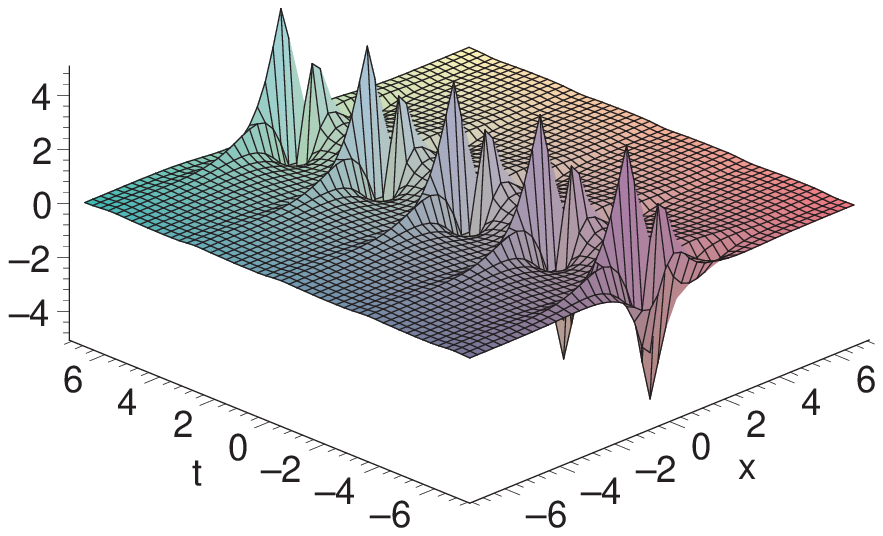}}}\qquad
$~~~~~~~~~~~~~$
{\rotatebox{0}{\includegraphics[width=5cm,height=4cm,angle=0]{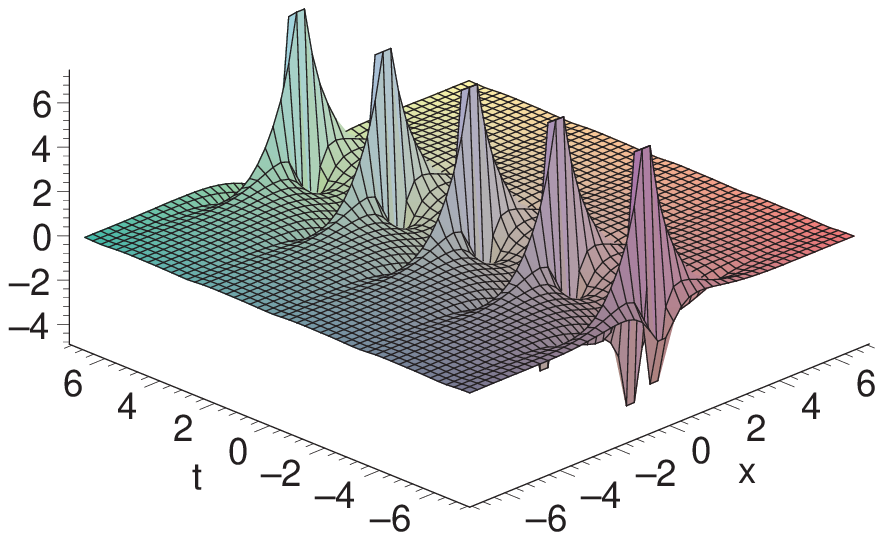}}}%\quad\\

$~~~~~~~~~~~~~~~~~~~~~~~~~~~~~~~~~~$$(a)$$
~~~~~~~~~~~~~~~~~~~~~~~~~~~~~~~~~~~~~~~~~~~~~~~~~~~~~~~~~~~~~~~~~~~~~~$
 $(b)$

 $~~~~$
{\rotatebox{0}{\includegraphics[width=5cm,height=4cm,angle=0]{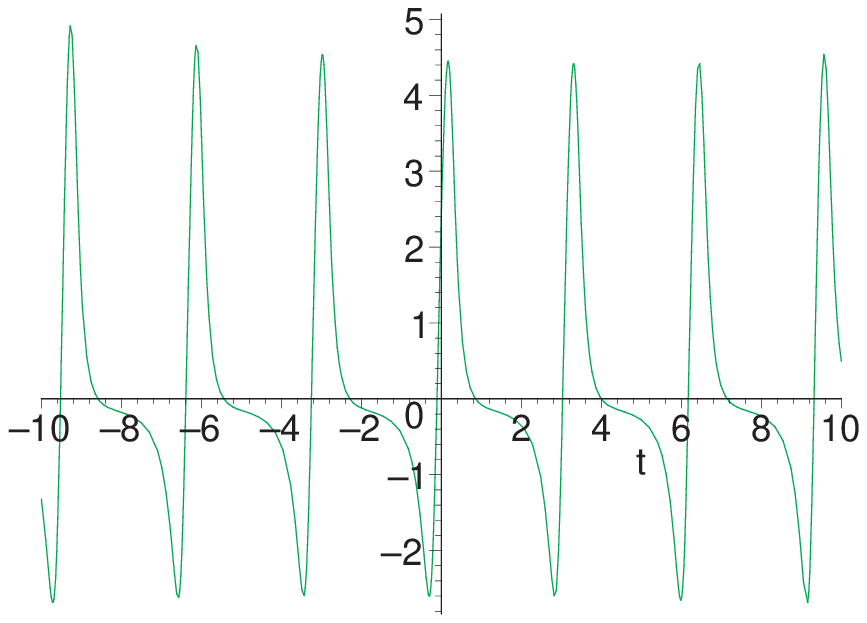}}}\qquad
$~~~~~~~~~~~~~$
{\rotatebox{0}{\includegraphics[width=5cm,height=4cm,angle=0]{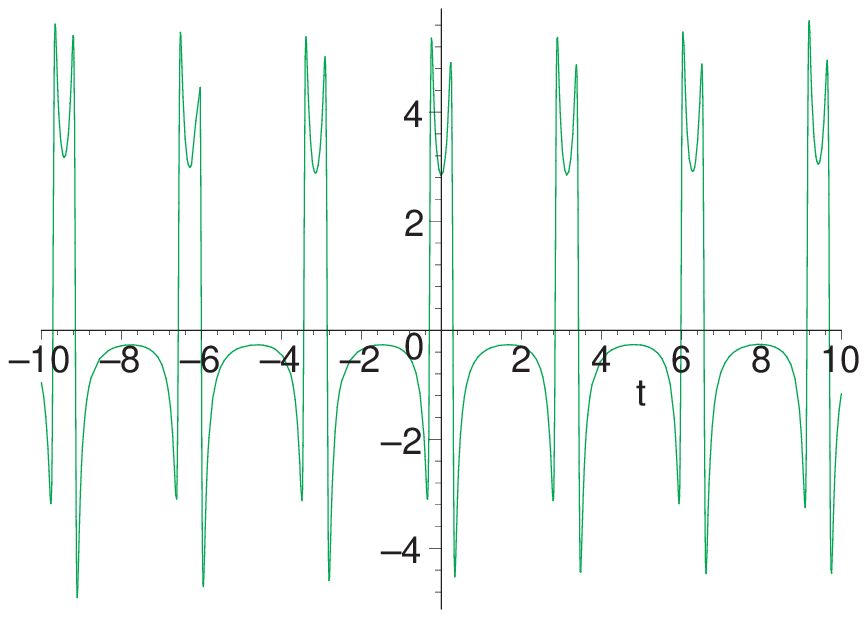}}}%\quad\\

$~~~~~~~~~~~~~~~~~~~~~~~~~~~~~~~~~~$$(c)$$
~~~~~~~~~~~~~~~~~~~~~~~~~~~~~~~~~~~~~~~~~~~~~~~~~~~~~~~~~~~~~~~~~~~~~~$
 $(d)$\\
\small{\textbf{Fig. 2.} (Color online) A  symmetric bright
two-soliton solution $|\psi(x,t)|^{2}$ of Eq.\eqref{1D NLSE} with
parameters: $\mu_{1}=0.1$, $\nu_{1}=0.1$, $\xiup_{1}=i$,
$\etaup_{1}=0.1i$, $\mu_{2}=1$, $\nu_{1}=1$, $\xiup_{2}=i$,
$\etaup_{2}=0.1i$, and $\epsilonup=0.01$. This figure shows that the
symmetric bright two-soliton periodic wave is spatially periodic in
two directions, but it is periodic in either the $x$ or $t$
directions. $(a)$ Perspective view for the real part of wave. $(b)$
Perspective view of the imaginary part wave. $(c)$ Wave propagation
pattern of the real part wave along the $t$ axis. $(d)$ Wave
propagation pattern of the imaginary part wave along the $t$ axis.}

\section{Bright  soliton interactions on the periodic
background} In this section, we investigate some bright soliton
interactions on the periodic background by using Eqs.\eqref{DT1} and
\eqref{DTN}. The simple exact solution of Eq.\eqref{1D NLSE} is the
plane wave
\begin{align}\label{initial seed solution}
&\psi(x,t)_{0}=u(x,t)_{0}e^{-i\frac{\epsilonup
x^{2}}{4}-\frac{\epsilonup t}{2}},\notag\\ &u(x,t)_{0}=\gamma
e^{i(kx+\omegaup t)},
\end{align}
where $\gamma$ and $k$ are real constants, and the frequency
$\omegaup$ solves the nonlinear dispersion relation
\begin{equation}
\omegaup=-k^{2}+\epsilonup
k+2\gamma^{2}-i\epsilonup+\frac{\epsilonup^{2}}{4}.
\end{equation}
Considering solution \eqref{initial seed solution} as the initial seed
solution of Eq.\eqref{1D NLSE}, we obtain the linear equations
\eqref{lax pair} and have the eigenfunction corresponding to the
eigenvalue $\lambda_{1}$ in the form
\begin{equation}\label{the eigenvalue}
\phiup[1,\lambda_{1}]_{1}=d_{1}e^{\mathscr{D}_{1}}+d_{2}e^{\mathscr{D}_{2}},\quad
\phiup[1,\lambda_{1}]_{2}=d_{3}e^{-\mathscr{D}_{1}}+d_{2}e^{-\mathscr{D}_{2}},
\end{equation}
with
\begin{align}
&\mathscr{D}_{1}=\frac{1}{2}i(kx+\omegaup t)+\Gamma(x+\Delta t),
\quad \mathscr{D}_{2}=\frac{1}{2}i(kx+\omegaup t)-\Gamma(x+\Delta
t),\notag\\
&d_{3}=\frac{d_{2}}{\gamma\sqrt{g_{0}}}\left(\frac{ik}{2}-\Gamma
-\lambda_{1}\right),\quad
\Gamma^{2}=\left(\lambda_{1}+\frac{k}{2}\right)^{2}-\gamma^{2},~~d_{4}=\frac{d_{1}}{\gamma\sqrt{g_{0}}}\left(\frac{ik}{2}+\Gamma
-\lambda_{1}\right),\notag\\
& \Delta=k\lambda_{1}-\epsilonup
k-2i\epsilonup\lambda_{1}-i\frac{\epsilonup
k}{2}-\frac{\epsilonup}{2}+i\gamma^{2}-i\frac{\epsilonup^{2}}{4}
+(2i\epsilonup-k)\left[\left(\lambda_{1}+\frac{k}{2}\right)^{2}-\gamma^{2}\right],
\end{align}
where $d_{1}$ and $d_{2}$ are two arbitrary complex constants.

Substituting expressions \eqref{the eigenvalue} into
formula \eqref{DTN}, the one-soliton and multi-soliton solutions on
the periodic background can be obtained with the iterative algorithm
of the Darboux transformation. The bright one-soliton solutions are
plotted in Figure 3, which shows four kinds of soliton profile
structures with different wave numbers. Although solitons in Figures
$3(c)$ and $3(d)$ hold larger wave numbers than those of Figures
$3(a)$ and $3(b)$, they both can propagate stably for long distances
by the results of numerical simulations of nonlinear pulses.\\

\noindent \textbf{Proposition 4.1.} \emph{Assuming that $
(\phiup[1,\lambda_{1}]_{1},\phiup[1,\lambda_{1}]_{2})^{T}$  be the
solution  of the linear eigenvalue problem \eqref{lax pair} with
eigenvalues $\lambda_{1}$, we obtain the bright two-soliton
solutions of Eqs.\eqref{1D NLSE} and \eqref{New 1D NLSE}
corresponding to eigenvalues $\lambda_{2}$ from formula \eqref{DTN}
with $N=2$
\begin{align} \label{DTNsolutions}
&\psi[2]=u[2]e^{-i\frac{\epsilonup x^{2}}{4}-\frac{\epsilonup
t}{2}},\notag\\
&u[2]=u(x,y)_{0}+2\sum_{i=1}^{2}\frac{(\lambda_{i}-\lambda_{i}^{*})\phiup[i,\lambda_{i}]_{1}
\phiup[i,\lambda_{i}]_{2}^{*}}{\phiup[i,\lambda_{i}]_{1}\phiup[i,\lambda_{i}]_{1}^{*}
+\phiup[i,\lambda_{i}]_{2}\phiup[i,\lambda_{i}]_{2}^{*}}.
\end{align}
The solution  of the linear eigenvalue problem \eqref{lax pair}
with eigenvalues $\lambda_{2}$ is
\begin{align}
&\phiup[2,\lambda_{2}]_{1}=\frac{(\lambda_{1}+\lambda_{1}^{*})\phiup[1,\lambda_{1}]_{1}
|\phiup[1,\lambda_{1}]_{2}|^{2}-\lambda_{1}^{*}\phiup[1,\lambda_{1}]_{1}
\phiup[1,\lambda_{1}]^{*}_{2}-\lambda_{1}\phiup[1,\lambda_{1}]_{1}\phiup[1,\lambda_{1}]^{2}_{2}}
{\phiup[1,\lambda_{1}]_{1}\phiup[1,\lambda_{1}]^{*}_{1}+\phiup[1,\lambda_{1}]_{2}\phiup[1,\lambda_{1}]^{*}_{2}},\notag\\
&\phiup[2,\lambda_{2}]_{2}=\frac{(\lambda_{1}-\lambda_{1}^{*})|\phiup[1,\lambda_{1}]_{1}|^{2}
\phiup[1,\lambda_{1}]_{2}+\lambda_{1}^{*}|\phiup[1,\lambda_{1}]_{1}|^{2}
\phiup[1,\lambda_{1}]^{*}_{2}-\lambda_{1}\phiup[1,\lambda_{1}]^{3}_{2}}
{\phiup[1,\lambda_{1}]_{1}\phiup[1,\lambda_{1}]^{*}_{1}+\phiup[1,\lambda_{1}]_{2}\phiup[1,\lambda_{1}]^{*}_{2}}.
\end{align}}

The bright two-soliton solutions are plotted in Figures 4.  Figure 4
depicts the bright two-soliton periodic wave solutions in the
three-dimensional space. From the interaction process in two sets of
figures, it can be clearly seen that the interacting solitons like
particles cross each other unaffectedly only by a phase shift, and
their respective amplitudes and velocities are the same as those
before collision. Further theoretical analysis for soliton solutions
shows that the sign and value of real part of the eigenvalue
determines the propagation direction of soltion and the amplitude,
respectively.\\

$~~~~~~~~$
{\rotatebox{0}{\includegraphics[width=5cm,height=4cm,angle=0]{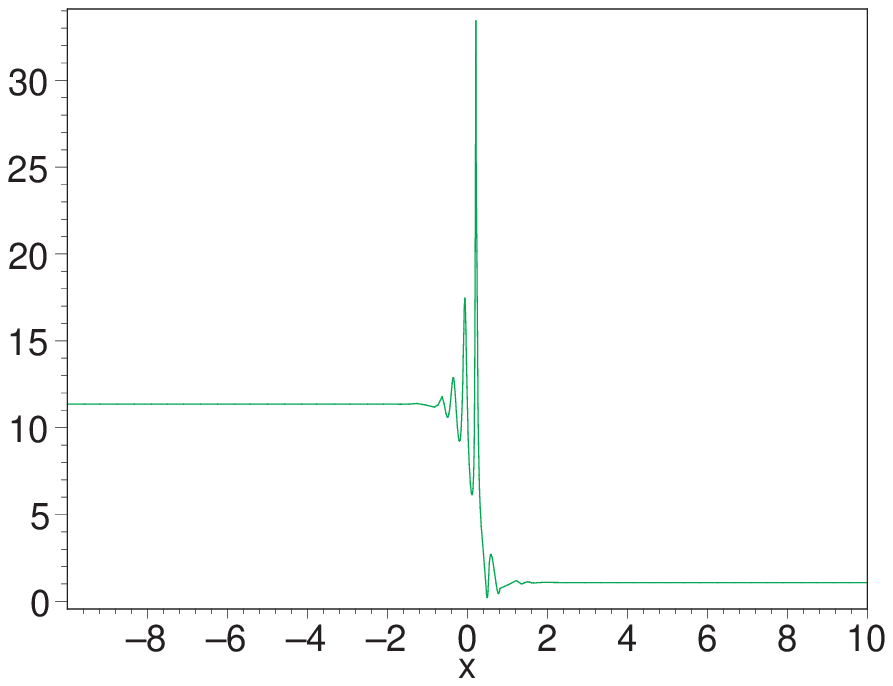}}}\qquad
$~~~~~~~~~~~~~$
{\rotatebox{0}{\includegraphics[width=5cm,height=4cm,angle=0]{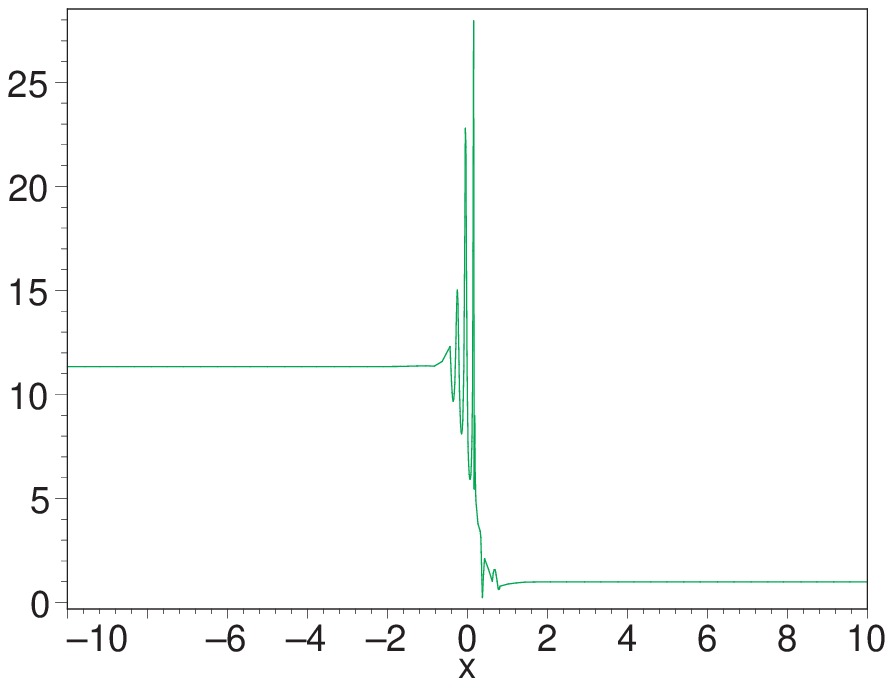}}}%\quad\\

$~~~~~~~~~~~~~~~~~~~~~~~~~~~~~~~~~~~~~$$(a)$$
~~~~~~~~~~~~~~~~~~~~~~~~~~~~~~~~~~~~~~~~~~~~~~~~~~~~~~~~~~~~~~~~~~~~~~~~$
 $(b)$

 $~~~~~~~~~~${\rotatebox{0}{\includegraphics[width=5cm,height=4cm,angle=0]{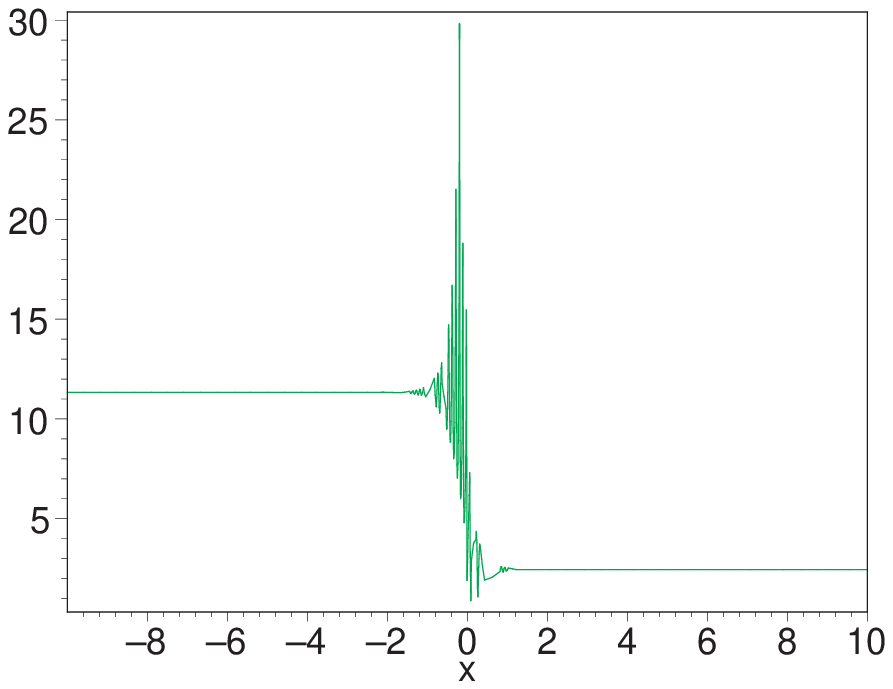}}}\qquad
$~~~~~~~~~~~~~$
{\rotatebox{0}{\includegraphics[width=5cm,height=4cm,angle=0]{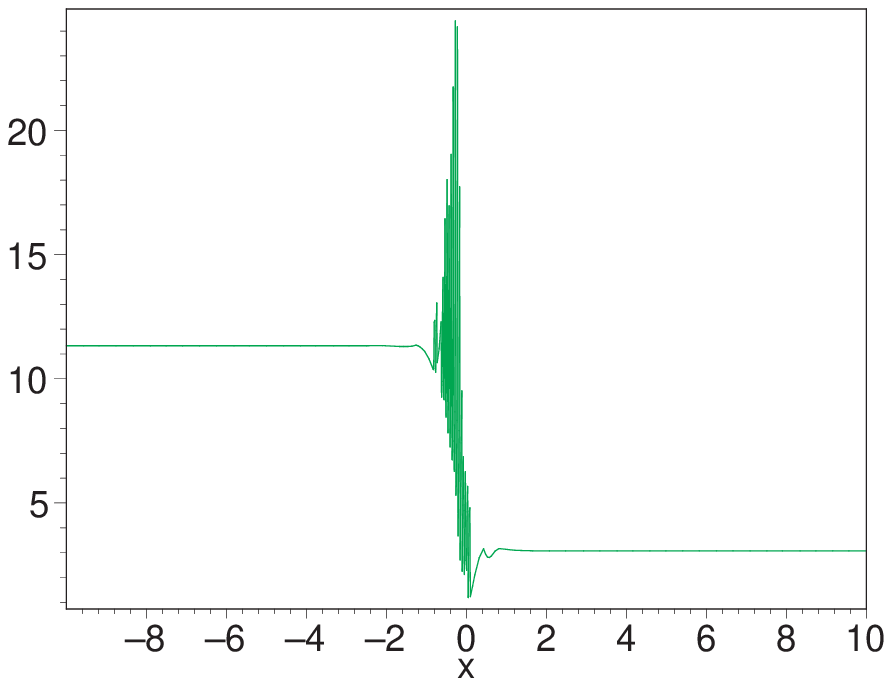}}}%\quad\\

$~~~~~~~~~~~~~~~~~~~~~~~~~~~~~~~~~~~~~$$(c)$$
~~~~~~~~~~~~~~~~~~~~~~~~~~~~~~~~~~~~~~~~~~~~~~~~~~~~~~~~~~~~~~~~~~~~~~~~~$
 $(d)$\\
\small{\textbf{Fig. 3.} (Color online) Four kinds of bright
one-soliton solutions $|\psi(x,t)|^{2}$ of Eq.\eqref{1D NLSE} with
different wave numbers $k$. The related physical quantities are
$\gamma=2, d_{1} =1, d_{2}=3, \lambda_{1}=5-2i, g_{0}=3,
\epsilonup=0.001$ and wave numbers $(a)$ $k=1$, $(b)$ $k=10$, $(c)$
$k=50$, $(d)$ $k=100.$}

$~~~~~~~~~~~~~~~~~~~~~~~~~~~~~~~~~~~~~~~~~~~~~~~~~$
{\rotatebox{0}{\includegraphics[width=5cm,height=4cm,angle=0]{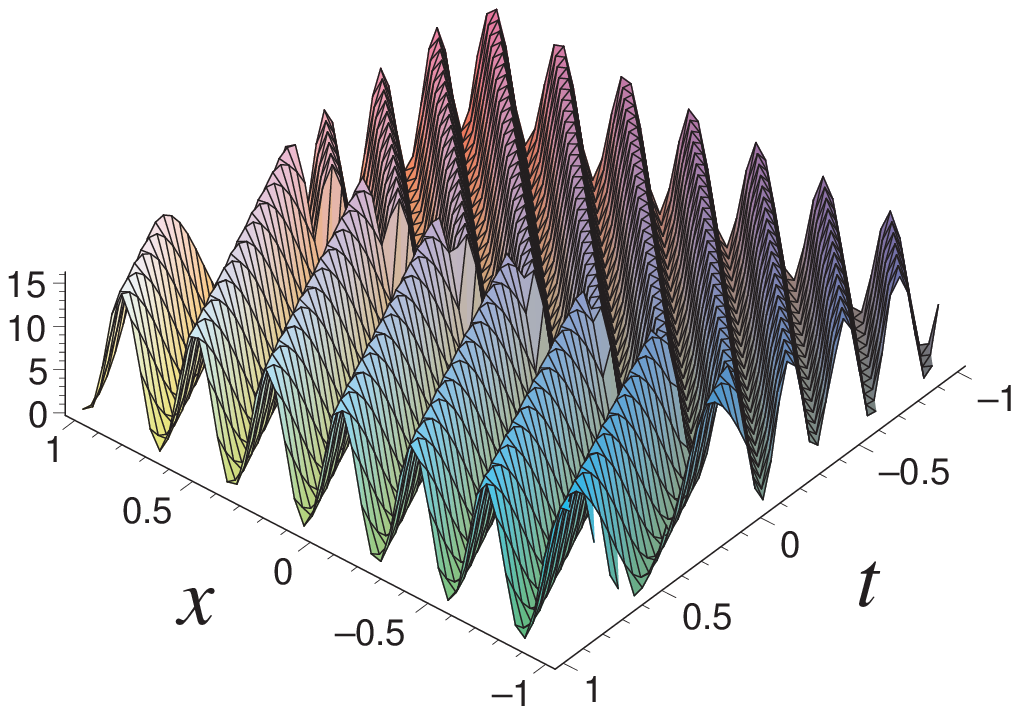}}}\qquad
\\

$~~~~~~~~~~~~~~~~~~~~~~~~~~~~~~~~~~~~~~~~~~~~~~~~~~~~~~~~~~~~~~~~~~~~~~~~~~~~~~$$(a)$\\

$~~~~~~~~~~~$
{\rotatebox{0}{\includegraphics[width=5cm,height=4cm,angle=0]{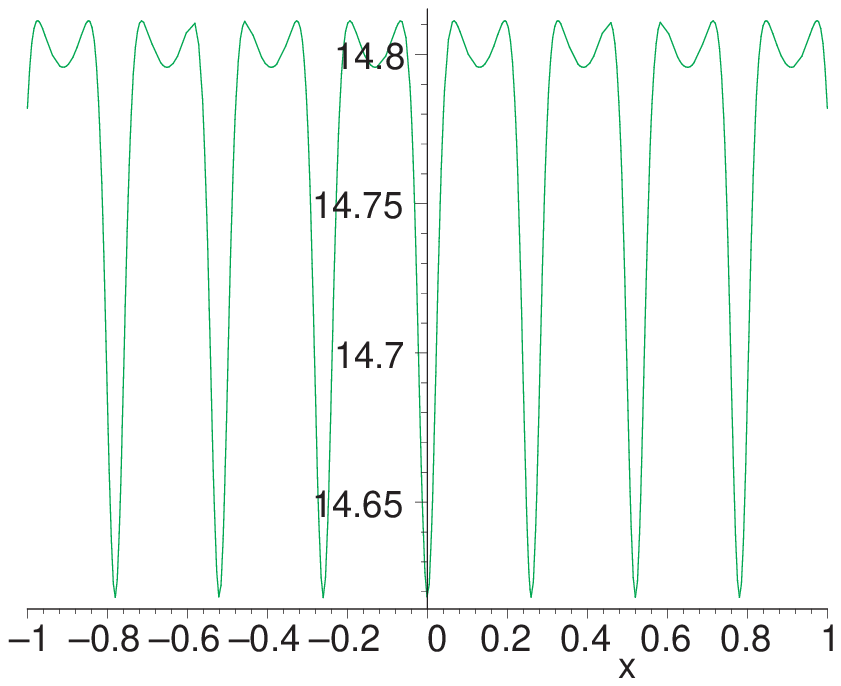}}}\qquad~~
$~~~~~~~~~~~~$
{\rotatebox{0}{\includegraphics[width=5cm,height=4cm,angle=0]{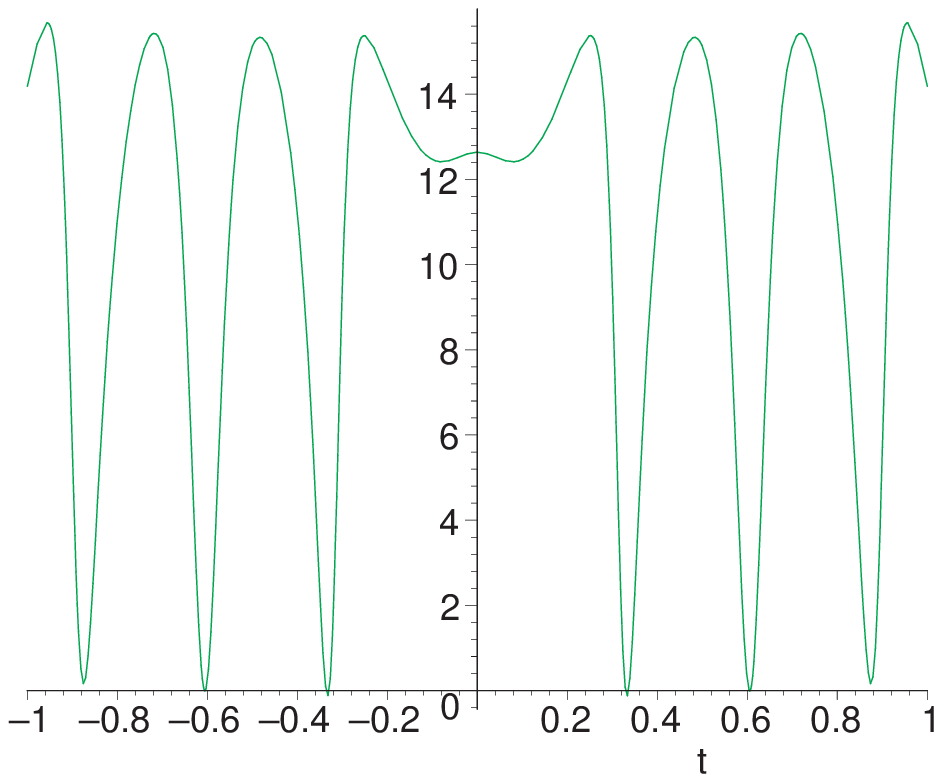}}}
%\quad\\

$~~~~~~~~~~~~~~~~~~~~~~~~~~~~~~~~~~~~~~$$(b)$$~~~
~~~~~~~~~~~~~~~~~~~~~~~~~~~~~~~~~~~~~~~~~~~~~~~~~~~~~~~~~~~~~~~~~~~~~~~~~~$
 $(c)$\\
\small{\textbf{Fig. 4.} (Color online) A  asymmetric bright
two-soliton solution $|\psi(x,t)|^{2}$ of Eq.\eqref{1D NLSE} with
parameters: $\gamma=2, d_{1}=1, d_{2}=3, \lambda_{1}=10+2i,
\lambda_{2}=10-2i, g_{0}=1, \epsilonup=0.001$ and wave numbers
$k=1$. This figure shows that the asymmetric bright two-soliton
periodic wave is spatially periodic in two directions, but it need
not be periodic in either the $x$ or $t$ directions. $(a)$
Perspective view of the wave.  $(b)$ Wave propagation pattern of the
wave along the $x$ axis. $(c)$ Wave propagation pattern of wave
along the $t$ axis.}

\section{Linear stability analysis}
In this section, we analyze the modulational instability of the
nonlinear plane waves. The phenomenon of modulational instability of
nonlinear waves has a significant importance in the theory and
experiment of nonlinear waves. The reason is that this phenomenon is
quite relevant applications in many different branches of physics
such as in condensate physics, plasma physics, hydrodynamics and
nonlinear optical fibers \cite{Kamchatnov}. Of particular interest
for studies of stability of the solution of the NLS equation is that
the envelope of a monochromatic plane wave propagates in nonlinear
medium in the presence of microwave noise pulses.

As we well-known that the exact plane wave solution for the focusing
NLS equation is to be linear unstable \cite{Wright}. According to the
method of investigating the linear stability of the NLS equation,
we perform a linear stability analysis of the nonlinear plane
waves solution for Eq.\eqref{1D NLSE}. The solution \eqref{initial
seed solution} is the plane wave with constant amplitude for
Eq.\eqref{1D NLSE}. We investigate the solution with the perturbation
in the form
\begin{align}\label{initial  perturbation seed solution}
&\psi(x,t)=u(x,t)e^{-i\frac{\epsilonup x^{2}}{4}-\frac{\epsilonup
t}{2}},\notag\\ &u(x,t)=u_{0}(x,t)\left[1+\varepsilonup
\tilde{u}(x,t)\right].
\end{align}
By substitution of
\eqref{initial perturbation seed solution} into Eq.\eqref{New 1D
NLSE},  the linearized disturbance equation at $o(\varepsilonup)$
becomes
\begin{equation}\label{disturbance equation}
i\tilde{u}_{t}+\tilde{u}_{xx}+i(2k-x\epsilonup)\tilde{u}_{x}+2\gamma^{2}a(t)e^{\epsilonup
t}(\tilde{u}^{*}+2\tilde{u})+\epsilonup
(x-1)\left(\frac{\epsilonup}{4}x+\frac{\epsilonup}{4}+k\right)\tilde{u}=0.
\end{equation}
By virtue of the linearity of Eq.\eqref{disturbance equation}, the
solution can be expressed as the following linear combination
\begin{equation}\label{linear combination}
\tilde{u}(x,t)=\mathscr{L}_{+}e^{i\theta(x-\Theta
t)}+\mathscr{L}^{*}_{-}e^{-i\theta(x-\Theta^{*} t)}.
\end{equation}
We characterize this solution by the real disturbance wave number
$\theta$ and the complex phase velocity $\Theta$. 
Substituting \eqref{linear combination} into \eqref{disturbance
equation} and collecting resonant terms, we have two linear
homogeneous equations
\begin{align}\label{two linear homogeneous equations}
&\left[\theta\Theta-\theta^{2}-\theta(2k-x\epsilonup)+4\gamma^{2}a(t)e^{\epsilon
t}+\epsilonup(x-1)\left(\frac{\epsilonup}{4}x+\frac{\epsilonup}{4}+k\right)\right]\mathscr{L}_{+}+
2\gamma^{2}a(t)e^{\epsilon t}\mathscr{L}_{-}=0,\notag\\
&2\gamma^{2}a(t)e^{\epsilon
t}\mathscr{L}_{+}+\left[-\theta\Theta-\theta^{2}+\theta(2k-x\epsilonup)+4\gamma^{2}a(t)e^{\epsilon
t}+\epsilonup(x-1)\left(\frac{\epsilonup}{4}x+\frac{\epsilonup}{4}+k\right)\right]\mathscr{L}_{-}=0.
\end{align}
By using the determinant of the matrix of coefficients of linear
Eqs.\eqref{two linear homogeneous equations}, the dispersion relation
for linearized disturbance can be expressed as
\begin{align}\label{5}
\Theta=&2k-x\epsilonup \pm
\frac{1}{\theta}\sqrt{\left(\frac{\epsilonup}{4}+k\right)^{2}\left[4\gamma^{2}a(t)e^{\epsilonup
t}+\theta^{2}-\epsilonup\left(\frac{\epsilonup}{4}+k\right)\right]^{2}-4\gamma^{4}a(t)^{2}e^{2\epsilonup
t}}\notag\\
=&2k-x\epsilonup \pm
\frac{1}{\theta}\sqrt{\left(\frac{\epsilonup}{4}+k\right)^{2}\left[4\gamma^{2}g_{0}e^{2\epsilonup
t}+\theta^{2}-\epsilonup\left(\frac{\epsilonup}{4}+k\right)\right]^{2}-4\gamma^{4}g_{0}^{2}e^{4\epsilonup
t}}~~(\rm{with}~~a(t)=g_{0}e^{\epsilonup t}),
\end{align}
where $0<\epsilonup\ll 1$ and $\gamma$, $k$ are real constants.

By virtue of the above relation \eqref{5}, if $\theta$ satisfies one of the
inequalities as follows
\begin{align}\label{ineq1}
&\theta^{2}\geq \frac{2\gamma^{2}g_{0}e^{2\epsilonup
t}}{\epsilonup/4+k}-4\gamma^{2}g_{0}e^{2\epsilonup t}+\epsilonup
\left(\frac{\epsilonup}{4}+k\right)=\left(\frac{2}{\epsilonup/4+k}-4\right)\gamma^{2}g_{0}e^{2\epsilonup
t}+\epsilonup \left(\frac{\epsilonup}{4}+k\right),\\\label{ineq2}
&\theta^{2}\leq -\frac{2\gamma^{2}g_{0}e^{2\epsilonup
t}}{\epsilonup/4+k}+4\gamma^{2}g_{0}e^{2\epsilonup t}-\epsilonup
\left(\frac{\epsilonup}{4}+k\right)=\left(4-\frac{2}{\epsilonup/4+k}\right)\gamma^{2}g_{0}e^{2\epsilonup
t}-\epsilonup \left(\frac{\epsilonup}{4}+k\right),
\end{align}
the frequency $\Theta$ is real at any value of the wavenumber
$\theta$, whereas $\Theta$ becomes complex.\\

\noindent\textbf{Proposition 5.1} \emph{(i) For Eq.\eqref{ineq1}, if
$\theta^{2}\geq \epsilonup \left(\frac{\epsilonup}{4}+k\right)$ and
$k\geq \frac{1}{4}>\frac{1}{2}-\frac{\epsilonup}{4}$
($0<\epsilonup\ll 1$), the frequency $\Theta$ is real at any value
of the time variable $t$. Otherwise, there is the instability region
of the modulational waves, and the disturbance will grow with time
exponentially. (ii) For Eq.\eqref{ineq2}, if $0\leq\theta^{2}\leq
-\epsilonup \left(\frac{\epsilonup}{4}+k\right)$ and
$k<-\frac{\epsilonup}{4}$ ($0<\epsilonup\ll 1$), the frequency
$\Theta$ is real at any value of the time variable $t$. Otherwise,
there is the instability region of the modulational waves, and the
disturbance will grow with time exponentially.}

Compared with the analytical method,
in the near future,  we intend to study the modulational instability numerically for the 1D NLSE \eqref{1D NLSE} since the linear stability
analysis actually shows that this instability is possble.

\section{ Conclusions and discussions}
In this paper, we investigate  a general form of nonlinear
Schr\"{o}dinger equation with time-dependent nonlinearity, which
describes  conservation laws, bright matter wave solitons and
modulational instability  in Bose-Einstein condensates with the
time-dependent interatomic interaction in an expulsive trapping
potential. By virtue of the Lax pair, we have further confirmed the
complete integrability of such a model by deriving an infinite
number of conservation laws. We have adopted the Darboux
transformation method to construct the multi-soliton solutions via
algebraic iterative algorithm. We have discussed four kinds of
bright one-soliton solutions $|\psi(x,t)|^{2}$  with different wave
numbers $k$, and two-periodic  bright solitons of Eq.\eqref{1D
NLSE} under the periodic background. Finally, we have analyzed the
linear stability of nonlinear plane waves in the presence of small
perturbation. In addition, it can be also applied to other spectral
problems in the field of nonlinear partial differential equation
appear in mathematical physics.

\section*{Acknowledgments}
The authors are grateful to  referees's very detailed comments and corrections for the first
version.
 The work is partially supported by the Doctoral Academic
Freshman Award of Ministry of Education of China under the grant 0213-812002, Doctaral Fund of
Ministry of Education of China under the grant 20100041120037, Natural Sciences
Foundation of China under the grant 11026165, 50909017 and the Fundamental Research Funds for the Central Universities
DUT11SX03.

%\end{CJK*}
\end{document}